\documentstyle[12pt]{article}

\def\be{\begin{equation}}
\def\ee{\end{equation}}
\def\ba{\begin{eqnarray}}
\def\ea{\end{eqnarray}}
\def\ga{\mathrel{\raise.3ex\hbox{$>$\kern-.75em\lower1ex\hbox{$\sim$}}}}
\def\la{\mathrel{\raise.3ex\hbox{$<$\kern-.75em\lower1ex\hbox{$\sim$}}}}

\begin{document}

\baselineskip=16pt 
\begin{titlepage} 
\rightline{UMN--TH--1844/00}
\rightline{TPI--MINN--00/10}
\rightline{hep-ph/0002229}
\rightline{February 2000}  
\begin{center}

\vspace{0.5cm}

\large {\bf Static Solutions for Brane Models with a Bulk Scalar Field}
\vspace*{5mm}
\normalsize

{\bf Panagiota Kanti, Keith A. Olive} and {\bf
Maxim Pospelov}

\smallskip 
\medskip 
 
{\it Theoretical Physics Institute, School of Physics and
Astronomy,\\  University of Minnesota, Minneapolis, MN 55455, USA} 

\smallskip 
\end{center} 
\vskip0.6in 
 
\centerline{\large\bf Abstract}

We present static solutions of the 5-dimensional Einstein 
equations in the brane-world scenario by using two different
approaches for the stabilization of the extra dimension.
Assuming a ``phenomenological'' stabilization mechanism, that
creates a non-vanishing $\hat T^5_5$ in the bulk, we construct a
two-brane model, which allows both branes to have positive
self-energies. We then consider a candidate theory for
the  dynamical stabilization, through the introduction of a massless 
scalar field in the bulk, which interacts 
with the branes. We find exact static solutions for the metric and
scalar  field in the bulk and 
demonstrate that the inter-brane distance is determined by the 
parameters of scalar field-brane interactions. However,
these solutions are always accompanied by a correlation
between the bulk cosmological constant, the brane self-energies
and the interaction terms of the scalar field with the branes
and thus cannot be considered as candidates for the phenomenologically 
viable stabilized geometry. We find that the aforementioned correlation
cannot be avoided even in the case of a single-brane solution 
with positive self-energy where the fifth dimension ends on a singularity.

\vspace*{2mm} 

\end{titlepage} 

\section{Introduction}

Two of the most serious problems which confronts unified theories today are
the hierarchy problem and the cosmological constant.  While supersymmetry
can stabilize the hierarchy \cite{hier}, the necessity to input mass scales
which differ by many orders of magnitude persists. In some respects, the
cosmological constant is even more severe, as many potential contributions
to the vacuum energy density must cancel to extremely high precision. 
It is quite plausible that the solution to both of these problems
lies beyond 4-dimensional field theory.  Indeed there have been several
recent attempts at attacking both of these problems in the context of higher
dimensional theories in the case of the hierarchy problem
\cite{RS,D1,ck,others} and in the case of the cosmological constant
\cite{occ,cc}.

In theories with extra dimensions (large or small), phenomenology and
cosmology must be restricted to a 3-brane solution in the larger theory. 
Indeed, considerable attention has been focused on one and two 3-brane
solutions. In particular, in the static solution of Randall and
Sundrum \cite{RS}, the scale factor $a$, is derived to be exponentially
decreasing as one moves away from a 3-brane with positive tension.
In a space-time with a compact extra dimension, a negative tension is
necessary, and a mass  hierarchy can be established between the two
branes.

In the absence of a stabilization mechanism for the modulus of the extra
dimension (radion), non-static solutions appeared problematic as the
cosmological expansion rate was found to depend on the energy density ($H
\simeq \rho$) rather than its square root as in the standard FRW Universe
\cite{LOW,LK,BDL}.  Solutions to this problem by adding both matter and
a cosmological constant on the two branes inevitably led to the wrong sign
for gravity on one of the branes \cite{NK,Berkl,Cline,Kor}.  

It was subsequently realized that the ``normal'' form of the Friedmann
equation is intimately related with the stabilization of the extra
dimension \cite{kkop,csak2,kkop2,kim}. Ideally, this should be
accomplished by a mechanism which works without any fine tuning of the
``input'' parameters and can be universally applied for any equation of
state on the brane. A consequence of such a stabilization is the
existence of (55)-component of the energy-momentum tensor in the bulk,
proportional to the trace of the energy-momentum tensor on the brane
\cite{kkop}. It was further shown \cite{kkop2} that this component arises
due to the shift of the minimum of the radion potential in response to the
presence of the brane. This way, the relation $T_{5}^{5}\simeq
-(2R)^{-1}(\rho -3p)$ arises naturally and is independent of the details of
the stabilization. See also Ref. \cite{E} for related constraints.

The apparent simplicity of the static
solution for the metric in the RS model is based on the exact fine-tuning of
the bulk and brane cosmological constants. The fine tuning is
exacerbated when perturbations of the brane self-energies with
matter densities are included. In this case, fine tuning between the energy
density and pressure components on the two different branes is needed. This
issue was readdressed in Ref. \cite{csak2}, where a phenomenological
stabilization potential for the transverse scale factor was introduced. 
The potential removes the need for the correlation between matter
densities on different branes. 

Given the necessity for a radion-fixing potential in any realistic
generalization, it is now fair to question the necessity of the negative
energy brane. Indeed, it should be possible to construct a solution for 
two positive self-energy branes if the distance between two branes
is stabilized. It was shown in \cite {kkop2,BDEL} that the general
solution to the Einstein's equations for the 3-space scale factor in the
presence of the negative bulk cosmological constant admits $cosh$-like
behaviour. For this solution, the usual 4D Friedmann equations for matter
trapped on a single brane can be easily obtained. However, the
cancellation of the effective cosmological constant on the brane is an
extra fine-tuning condition. Because of the minimum in $a$, the same
$cosh$-like solutions should be able to accommodate two positive self-energy
branes placed on  opposite sides of the minimum. Here we plan to study
such static two-brane configurations with positive self-energies.  We will 
determine the allowed values of the parameters in this model and investigate
the possible hierarchy between scale factors on two different branes.

Irrespective of the size of the extra dimension, it is natural to
expect that the brane self-energy is large, on the order of the fifth
power of the fundamental 5-dimensional Plank scale. On the other hand,
the matter density, $\rho $, is small in these units no matter how low
the fundamental scale might be. This is true even in the extreme case
when $M_{5}\sim 1$ TeV, $\rho \ll $ TeV$^{4}$. As such, it
is clear that a natural mechanism for the cancellation of the
effective cosmological constant on the brane is another very important
question which has to be resolved in order to connect the
brane-world proposal to reality. To this end, we first study static
solutions to Einstein equations and neglect the matter density $\rho
$. The time independence of these solutions
automatically means that the effective cosmological constant on the brane is
equal to zero. If such solutions are found, one can then
perturb them by including a small $\rho $ in order to get a consistent
phenomenological and cosmological description.

The stabilization of the extra dimension with a bulk scalar field was
discussed by Goldberger and Wise in Refs. \cite{GW}. See also Ref.
\cite{tanaka}. There, the original RS solution was modified by including
a scalar field in the bulk, which has an interaction (potential) with
the two branes. This stabilization does not evade the fine tuning,
which in Goldberger-Wise approach is the same fine tuning as in Refs.
{\cite{RS,RS2}}, that is the fine tuning between brane self-energies and
bulk cosmological constant. It is important, however, to study this
mechanism in more detail in order to understand to what extent it
depends on the specific assumptions concerning the scalar field, its
potential in the bulk and interactions with the branes.

The purpose of this letter is two-fold. First, we derive static
solutions to Einstein's equations with two branes with positive
self-energies by allowing the value of $T_{55}$ to be non-zero in the
bulk. We find that such a solution can accommodate any 
positive values of the brane self-energies between zero and a limiting value 
corresponding to the brane self-energy in the Randall-Sundrum model. 
The ratio of the scale factors on the two branes is determined through
the deviations of the brane self-energies from this limiting value.
Secondly, we find exact static solutions to the Einstein's equations
in the presence of a massless scalar field, with the bulk
energy-momentum tensor given {\em only} by a cosmological constant and
the energy-momentum tensor of this field. We argue that in this case
the proper stabilization of the extra dimension and/or cancellation of
the effective cosmological constant on the brane is not possible unless
some specific fine tuning conditions are satisfied. Finally, we
present the single-brane configuration with the spacetime ending on
a true singularity in the extra dimension and comment on the subject
of fine tuning in this case. This solution generalizes the Randall-Sundrum
model and shows that the exponentially decaying scale factor 
eventually ends on the singularity, situated at the point in the extra 
dimension determined through the strength of the brane-scalar field
interaction.

\section{Static two-brane models with phenomenological
stabilization of extra dimension}

We start with the description of the geometrical framework of our analysis.
The line-element of the 5-dimensional spacetime is given by the following
ansatz
\begin{equation}
ds^2=a^2(y)\,(-dt^2 + \delta_{ij} dx^i dx^j) + b^2(y) dy^2\,, 
\label{metric}
\end{equation}
where $\{t,x^i\}$ and $y$ denote the usual, 4-dimensional spacetime and the
extra dimension, respectively. Here, we focus only on static configurations
of the spacetime background and ignore any time dependence of the conformal
factor $a$ and the scale factor $b$ along the extra dimension. Without
loss of generality, we can assume $b=1$.

We will also assume that the two 3-branes with positive self-energies
$\Lambda_1$ and $\Lambda_2$ are located at 
$y=y_1$ and
$y=-y_2$, respectively.  In the region
between the two 3-branes, a non-vanishing cosmological constant $\Lambda_B$
is assumed to exist. The action functional that describes the above
(4+1)-dimensional, gravitational theory has the following form
\be
S=-\int d^{4}x\,dy\,\sqrt{-\hat{g}}\,\biggl\{\frac{
M_{5}^{3}}{16\pi }\,\hat{R}+\Lambda _{B}+\Lambda _{1}\,\delta
(y-y_{1})+\Lambda _{2}\,\delta (y+y_{2})\biggr\}\,.
\label{action0}
\ee
In the above, $M_{5}$ is the fundamental 5-dimensional Planck mass and
the hat denotes 5-dimensional quantities.  The existence of some
stabilization mechanism is also assumed which ensures that the distance
between the two branes remains fixed. According to Refs.
\cite{kkop,csak2,kkop2,kim}, this requires a bulk value for
$\hat{T}_{5}^{5}$ different from $-\Lambda _{B}$. In this sense, the
solutions that we derive here, are
generalizations of the Randall-Sundrum constructions which allow the
existence of a non-trivial bulk value of $\hat T_{5}^{5}$ and,
consequently, positive self-energies for both branes. 

The variation of the action (\ref{action0}) with respect to
the 5-dimensional metric tensor $\hat g_{MN}$ leads to Einstein's equations,
which for the spacetime background (\ref{metric}) take the form
\begin{eqnarray}
\hat{G}_{00} &=&-3a^{2}\Biggl\{\frac{a^{\prime \prime }}{a}+\biggl(\frac{
a^{\prime }}{a}\biggr)^{2}\Biggr\} =\hat{\kappa}^{2}\,\hat{T}_{00}\,,
\label{001} \\[4mm]
\hat{G}_{ii} &=&3a^{2}\Biggl\{\frac{a^{\prime \prime }}{a}+\biggl(\frac{
a^{\prime }}{a}\biggr)^{2}\Biggr\} =\hat{\kappa}^{2}\,\hat{T}_{ii}\,,
\label{ii1} \\[4mm]
\hat{G}_{55} &=&6\,\left( \frac{a^{\prime }}{a}\right) ^{2}=\hat{\kappa}
^{2}\,\hat{T}_{55}\,,  \label{551}
\end{eqnarray}
where $\hat{\kappa}^{2}=8\pi G_{N}^{(5)}=8\pi /M_{5}^{3}$ and the primes
denote differentiation with respect to $y$. Note that the (05)-component of
Einstein's equations vanishes identically due to the time-independence of
the line-element (\ref{metric}).

Taking into account the contributions from the bulk cosmological constant
and the brane self-energies, the energy-momentum tensor that appears on
the rhs of Einstein's equations can be written as
\begin{equation}
\hat{T}_{\,\,\,\,\,N}^{M}=\Bigr[\Lambda _{B}+\Lambda _{1}\,\delta
(y-y_{1})+\Lambda _{2}\,\delta (y+y_{2})\Bigr]\,(-\delta_{\,\,\,\,\,N}^{M})\,.
\label{total}
\end{equation}
In addition, we allow the (55)-component to deviate from this
form due to the  existence of radius stabilization potential
\cite{kkop2}.  It is straightforward to see that, for the above choice,
eqs. (\ref{001}) and (\ref{ii1}) reduce to the same differential
equation for the conformal factor $a(y)$. In the bulk, this can be
conveniently rewritten as
\be
(a^{2})^{\,\prime \prime }=\frac{2\hat{\kappa}^{2}}{3}(-\Lambda _{B})\,a^{2}\,. 
\ee
In the case of a negative bulk cosmological constant, $\Lambda_B<0$, the 
general solution for $a^{2}(y)$ in the bulk is given by an arbitrary linear
combination of rising and falling exponents, 
\begin{equation}
a^{2}(y)=A\exp\left( {-\sqrt{\frac{2\hat{\kappa}^{2}}{3}|\Lambda _{B}|}y}
\right)+B\exp\left(
{\sqrt{\frac{2\hat{\kappa}^{2}}{3}|\Lambda _{B}|}y}\right)\,,
\end{equation}
where $A$ and $B$ are integration constants. The conformal factor must
also satisfy boundary conditions at
$y=y_1$ and $y=-y_2$ which depend on the brane self-energies,
$\Lambda _{1}$ and $\Lambda _{2}$. It is clear that 
$\Lambda _{1,2} > 0$ can be accommodated only if the solution for
$a^{2}(y)$ in the bulk, is not monotonic. In this case,
$a^{2}(y)$ is given by a hyperbolic cosine, and without a loss of
generality we can place the minimum of this function, $y_{0}$, at the
point $y=0$ and redefine
$y_{1}$ and $-y_{2}$ accordingly. Then, the solution for the conformal
factor takes the form
\begin{equation}
a^{2}(y)= a_0^2\, \cosh\left( \sqrt{\frac{2\hat{\kappa}^{2}}{3}
|\Lambda _{B}|}y\right)\,.
\label{cosh}
\end{equation}

The embedding of the two 3-branes with zero thickness in the 5-dimensional
manifold creates a discontinuity of the first derivative, with respect to $y$,
of the conformal factor $a(y)$. This, in turn, leads to the appearance of a
Dirac delta function in the (00) and (ii)-components of Einstein's equations
(\ref{001})-(\ref{ii1}), where its second derivative appears. By matching
the coefficients of the delta functions in the aforementioned equations,
the following {\it jump} conditions at the points $y=y_{1}$ and $y=-y_{2}$
emerge 
\be
\frac{[a^{\prime }]_{1}}{a_{1}}=-\frac{\hat{\kappa}^{2}}{3}\,\Lambda _{1}\,,
\qquad
\frac{[a^{\prime }]_{2}}{a_{2}}=-\frac{\hat{\kappa}^{2}}{3}\,\Lambda _{2}\,,
\label{jump0} 
\ee
where the subscripts $1$ and $2$ denote quantities evaluated at 
$y=y_1$ and $y=-y_2$, respectively. Using the expression (\ref{cosh}),
the above conditions can be rewritten as
\begin{eqnarray}
\tanh\left( \sqrt{\frac{2\hat{\kappa}^{2}}{3}|\Lambda _{B}|}\,y_{1}\right) &=&
\frac{\Lambda _{1}}{\sqrt{6|\Lambda
_{B}|/\hat{\kappa}^{2}}} \equiv \frac{\Lambda _{1}}{\Lambda _{RS}}\,, 
\label{bc1} \\ 
\tanh\left( \sqrt{\frac{2\hat{\kappa}^{2}}{3}|\Lambda
_{B}|}\,y_{2}\right) &=&\frac{\Lambda _{2}}{\sqrt{6|\Lambda
_{B}|/\hat{\kappa}^{2}}}=\frac{
\Lambda _{2}}{\Lambda _{RS}}\,.  \label{bc2}
\end{eqnarray}
As one can see, the
position of the branes is determined by their self-energies. Moreover,
these two conditions show that the static solution (\ref{cosh}) can
arise only if 
\begin{equation}
0\leq \Lambda _{1},\Lambda _{2}\leq \Lambda _{RS}\,.
\end{equation}

The limiting case of $\Lambda _{i}=\Lambda _{RS}$ corresponds to $
y_{i}\rightarrow \infty $ and effectively reproduces the solution of Ref. 
\cite{RS2} with the exponentially decaying conformal factor. The ratio of scale
factors on the two branes can be expressed in terms of ``detuning'' of $
\Lambda _{i}$ from the limiting values $\Lambda _{RS}$,
\begin{equation}
\frac{a_{2}^{2}}{a_{1}^{2}}=\sqrt{\frac{\Lambda _{RS}^{2}-\Lambda _{1}^{2}}
{\Lambda _{RS}^{2}-\Lambda _{2}^{2}}}\,.
\end{equation}
In principle, this ratio can be very large or very small, depending on the
relative size of these detunings. In order to solve the hierarchy problem,
we must assume that the observable matter fields are localized to the brane
with the smaller scale factor. Thus, we have demonstrated that the
gauge hierarchy problem can be resolved by  a ``geometrical" explanation
\`a la Ref. \cite{RS} with two positive self-energy branes.

Clearly, the above solution cannot arise without a
contribution to the (55)-component of the
energy-momentum tensor, other than $-\Lambda_{B}$. The value of $\hat
T_{55}$ consistent with the solution (\ref{cosh}) can be easily
determined by substituting the solution for the conformal factor in eq.
(\ref{551}), and is found to be 
\begin{equation}
\hat T_{55}=|\Lambda _{B}|-\frac{|\Lambda _{B}|}{{\rm cosh}^{2}
\left(\sqrt{\frac{2\hat{\kappa}^{2}}{3}|\Lambda _{B}|}\,y\right) }\,.
\end{equation}
If the inter-brane distance, $y_1+y_2$, is large as compared to the length
scale given by $1/\sqrt{\frac{2\hat{\kappa}^{2}}{3}|\Lambda _{B}|}$, 
the (55)-component of the energy-momentum tensor deviates from $|\Lambda_B|$ 
only in the vicinity of $y=0$ near the minimum of the scale factor.
Using eqs. (\ref{bc1}) and (\ref{bc2}), we can rewrite this expression 
in the following form,
\ba
\hat T_{55}=|\Lambda _{B}|\left(1-\frac{a_i^4}{a^4}\left[1-\frac{\Lambda_i^2}
{\Lambda_{RS}^2}\right]\right)= |\Lambda _{B}|- \frac{a_i^4}{a^4}
\frac{\Lambda_i^2\hat{\kappa}^{2}}{6\,{\rm sinh}^2
\left(\sqrt{\frac{2\hat{\kappa}^{2}}{3}|\Lambda _{B}|}\,y_i\right) },
\ea
which coincides with the expression obtained in Ref. \cite{kkop2}.
$a_i, \Lambda_i$, and $y_i$ can be evaluated on {\em either}
brane.

We should stress at this point that the distance between the two branes
(or, equivalently, the volume of the extra dimension) turns out to be fixed
in terms of the brane self-energies and the bulk cosmological constant.
In the limit of small bulk cosmological constant, the relations (\ref{bc1})
and (\ref{bc2}) take a remarkably simple form and can be combined to
give the result
\begin{equation}
\Lambda _{1}+\Lambda _{2}+2(y_{1}+y_{2})\Lambda _{B}=0.
\end{equation}
This is nothing other than the condition of mutual cancellation between
bulk and brane contributions to the effective cosmological constant,
and, as such, is an extra fine-tuning condition which the radius
stabilization has to satisfy. When we treat this stabilization
``phenomenologically'', by introducing a (55)-component for the
energy-momentum tensor in the bulk, the mechanisms which could ensure
this cancellation, simply cannot be addressed. Thus, we proceed to the
considerations of 5D gravity plus a scalar field in the bulk
interacting differently with the two branes, which was proposed in 
\cite{GW} to be a viable {\em dynamical} stabilization mechanism.

\section{Gravity and a massless scalar in extra dimensions}

In this section, we assume the existence of a scalar field, $\hat{\phi}$,
in the bulk, in addition to a bulk cosmological constant
$\Lambda_B$. We, now, choose the two 3-branes to be located at the 
points $y=0$ and $y=L$. The bulk scalar field is minimally coupled to gravity but may
have different interactions with the two branes. The action functional
of the theory, now, takes the form
\begin{eqnarray}
&~&\hspace*{-1cm}S=-\int d^{4}x\,dy\,\sqrt{-\hat{g}}\,\Biggl\{\frac{
M_{5}^{3}}{16\pi }\,\hat{R}+\Lambda _{B}+\frac{1}{2}\,\partial _{M}\hat{\phi}
\,\partial ^{M}\hat{\phi}+V_B(\hat{\phi})  \nonumber \\[3mm]
&~&\hspace*{4cm}+\Bigl[\Lambda _{1}+V_{1}(\hat{\phi})\Bigr]\,\delta (y)+\Bigl
[\Lambda _{2}+V_{2}(\hat{\phi})\Bigr]\,\delta (y-L)\Biggr\}\,.
\label{action}
\end{eqnarray}
In the above expression,
$V_B$, $V_{1}$, and $V_{2}$ are the bulk potential and the brane interactions
of the scalar field on the brane 1 and 2, respectively. As before,
$\Lambda _{1}$ and $\Lambda _{2}$ are the constant self-energies
of the two branes. Non-vanishing bulk potentials were also considered
in \cite{scalar}.

In the presence of the bulk scalar field, the Einstein's equations
(\ref{001})-(\ref{551}) are supplemented by the equation of motion for
the scalar field, which has the form 
\begin{equation}
\frac{1}{a^{4}}\,\frac{d}{dy}\left( a^{4}\,\frac{d\hat{\phi}}{dy}\right) =
\frac{\partial V_B(\hat{\phi})}{\partial \hat{\phi}}+\frac{\partial
V_{1}(\hat{
\phi})}{\partial \hat{\phi}}\,\delta (y)+\frac{\partial V_{2}(\hat{\phi})}{
\partial \hat{\phi}}\,\delta (y-L)\,.  \label{scalar}
\end{equation}

The energy-momentum tensor of the theory is also modified compared to the
expression (\ref{total}) of the previous section. The interaction terms
of the scalar field on the two branes, $V_1$ and $V_2$, will contribute to
the total brane self-energies while the bulk energy-momentum tensor 
may now be written as
\begin{equation}
\hat{T}_{\,\,\,\,\,N}^{M}=-\Lambda_B\,\delta_{\,\,\,\,\,N}^{M}
+ \hat T_{\,\,\,\,\,N}^{M}(\hat \phi)\,,
\label{totalbulk}
\end{equation}
where
\begin{equation}
\hat{T}_{MN}(\hat{\phi})=\partial _{M}\hat{\phi}\,\partial _{N}\hat{\phi}-
\hat{g}_{MN}\,\biggl[\frac{1}{2}\,\partial _{P}\hat{\phi}\,\partial ^{P}
\hat{\phi}+V_B(\hat{\phi})\biggl]\,.
\end{equation}

Let us, first, concentrate on the equation of motion of the scalar field in the
bulk where the last two terms on the rhs of eq. (\ref{scalar}) vanish. In
order to understand to which extent the proposed stabilization of the extra
dimension with the scalar field \cite{GW} depends on the specific
assumptions about its self-interaction, we take the potential in the bulk to
be identically zero. We also notice that the choice $V_B(\hat{\phi})=0$
allows us to easily integrate the lhs of eq. (\ref{scalar}) 
with respect to $y$ and find $\hat{\phi}
^{\prime }$ in terms of the conformal factor $a(y)$. This, in turn, will
lead to the determination of the conformal factor in the presence of the
scalar field in the bulk, i.e. the backreaction of the scalar field on the
spacetime geometry. Integrating eq. (\ref{scalar}), we obtain the result 
\begin{equation}
\hat{\phi}^{\prime }(y)=\frac{c\,a_{0}^{4}}{a^{4}(y)}\,,  \label{first}
\end{equation}
where $c$ is a constant and $a_{0}=a(y=0)$. When the above expression
is combined with Einstein's equations (\ref{001})-(\ref{551}) and the
expression for the energy-momentum tensor in the bulk (\ref{totalbulk}),
we are led to the following system of differential equations for $a(y)$ 
\begin{eqnarray}
\frac{a^{\prime \prime }}{a}+\biggl(\frac{a^{\prime }}{a}\biggr)^{2} &=&
\frac{\hat{\kappa}^{2}}{3}\biggl(-\Lambda _{B}-\frac{c^{2}a_{0}^{8}}{2a^{8}}
\biggr)\,,  \label{fin1} \\[4mm]
2\,\biggl(\frac{a^{\prime }}{a}\biggr)^{2} &=&\frac{\hat{\kappa}^{2}}{3}
\biggl(-\Lambda _{B}+\frac{c^{2}a_{0}^{8}}{2a^{8}}\biggr)\,.  \label{fin2}
\end{eqnarray}
Rearranging the above two equations, we are led to a single differential
equation
\begin{equation}
\frac{(a^{4})^{\prime \prime }}{4a^{4}}=-\frac{2\hat{\kappa}^{2}}{3} 
\,\Lambda _{B}\,,  \label{basic}
\end{equation}
which can be easily integrated to give the solution for the conformal factor 
$a(y)$ in terms of the bulk cosmological constant. The substitution of the
solution in any of the original equations (\ref{fin1})-(\ref{fin2}) and the
boundary condition $a(y=0)\equiv a_{0}$ will determine any arbitrary
integration constants. In this way, we obtain the following solution 
\begin{equation}
a^{4}(y)=a_{0}^{4}\,\frac{|y-y_{0}|}{y_{0}}\,,\qquad y_{0}=\sqrt{\frac{3}{4 
\hat{\kappa}^{2}c^{2}}}\,,  \label{solution}
\end{equation}
in the case of vanishing $\Lambda _{B}$, which is similar to the solution
found in \cite{cc}, and
\begin{equation}
a^{4}(y)=a_{0}^{4}\,\frac{\sin (\omega |y-y_{0}|)}{\sin (\omega y_{0})} 
\,,\qquad y_{0}=\frac{1}{\omega }\,Arc\sin \sqrt{\frac{2\Lambda _{B}}{c^{2}}} 
\,,  \label{solpos}
\end{equation}
or 
\begin{equation}
a^{4}(y)=a_{0}^{4}\,\frac{\sinh (\omega |y-y_{0}|)}{\sinh (\omega y_{0})} 
\,,\qquad y_{0}=\frac{1}{\omega }\,Arc\sinh \sqrt{\frac{2|\Lambda _{B}|}{ 
c^{2}}}\,,  \label{solneg}
\end{equation}
for positive or negative, respectively, $\Lambda _{B}$. The parameter $ 
\omega $ appearing in the above expressions is defined as 
\begin{equation}
\omega ^{2}=\frac{8\hat{\kappa}^{2}}{3}\,|\Lambda _{B}|\,.  \label{def}
\end{equation}
Note that all the above solutions are characterized by the existence of
a spacetime singularity at $y=y_0$, where the conformal factor vanishes
while the first derivative of the scalar field (\ref{first}) diverges.
By placing a second brane at a point
$y=L<y_{0}$, we can ensure that the above solutions are well defined
everywhere. Hereafter, we concentrate on the case of a negative bulk
cosmological constant, however, similar conclusions can be drawn in the
other two cases as well.

The inhomogeneity in the distribution of matter in the 5-dimensional
manifold leads to a discontinuity of the first derivative, with respect to $y$,
not only of the conformal factor $a(y)$, but of the bulk scalar field
$\hat{\phi}(y)$, too. By following the same method as in section 2, i.e.
by matching the coefficients of the delta functions in the equations where
their second derivatives appear, the following {\it jump} conditions,
for both the conformal factor and the scalar field, emerge
\begin{eqnarray}
&~&\frac{[a^{\prime }]_{0}}{a_{0}}=-\frac{\hat{\kappa}^{2}}{3}\,\Bigl 
[\Lambda _{1}+V_{1}(\hat{\phi}_{0})\Bigr]\,,\qquad [\hat{\phi}^{\prime
}]_{0}=\frac{\partial V_{1}(\hat{\phi})}{\partial \hat{\phi}}\Biggl|_{y=0}\,,
\label{jumpa} \\[4mm]
&~&\frac{[a^{\prime }]_{L}}{a_{L}}=-\frac{\hat{\kappa}^{2}}{3}\,\Bigl 
[\Lambda _{2}+V_{2}(\hat{\phi}_{L})\Bigr]\,,\qquad [\hat{\phi}^{\prime
}]_{L}=\frac{\partial V_{2}(\hat{\phi})}{\partial \hat{\phi}}\Biggl|_{y=L}\,,
\label{jumpb}
\end{eqnarray}
where the subscripts $0$ and $L$ denote quantities evaluated at $y=0$ and $ 
y=L$, respectively. In the
above, we have used the fact that the energy-momentum tensor on the two
branes is generated by the interaction terms of the bulk scalar field and
the brane self-energies. By using the expressions (\ref{first}) and (\ref
{solneg}), for the first derivative of the scalar field and the solution
for the conformal factor in the bulk, respectively, the above conditions may
be written as
\begin{eqnarray}
&~&\omega \,\coth (\omega y_{0})=\frac{2\hat{\kappa}^{2}}{3}\,\Bigl[\Lambda
_{1}+V_{1}(\hat{\phi}_{0})\Bigr]\,,\qquad 2c=\frac{\partial V_{1}(\hat{\phi})
}{\partial \hat{\phi}}\Biggl|_{y=0}\,,  \label{jump1} \\[4mm]
&~&\omega \,\coth [\omega \,(y_{0}-L)\Bigr]=-\frac{2\hat{\kappa}^{2}}{3}\,
\Bigl[\Lambda _{2}+V_{2}(\hat{\phi}_{L})\Bigr]\,,\qquad 2c\,\frac{a_{0}^{4}}{
a_{L}^{4}}=-\frac{\partial V_{2}(\hat{\phi})}{\partial \hat{\phi}}\Biggl|
_{y=L}\,.  \label{jump2}
\end{eqnarray}

A close examination of the above equations renders the allowed values for
the brane self-energies, ``dressed'' with the interaction with $\hat{\phi}$.
Assuming that the positive self-energy brane is situated at $y=0$,
we arrive at the following allowed ranges for the effective self-energies:
\begin{eqnarray}
\Lambda _{RS} &\leq &\Lambda _{1}+V_{1}(\hat{\phi}_{0})\leq \infty\,,  \\[3mm]
-\infty  &\leq &\Lambda _{2}+V_{2}(\hat{\phi}_{L})\leq -\Lambda _{RS}\,,
\end{eqnarray}
from which we immediately conclude that this solution cannot accommodate two
positive self-energy branes. 
Of course, we can choose both
$\Lambda _{1}$ and $\Lambda _{2}$ to be positive and remain
consistent with the boundary conditions (\ref{jump1})-(\ref{jump2})
provided that the potential on one of the branes is negative making the
``dressed'' brane self-energy negative, i.e. $\Lambda
_{2}+V_{2}(\hat{\phi}_{L})<0$. 
The fact that one of the two branes has a negative total energy density
follows from the form of the solution (\ref{solneg}) for the conformal
factor $a(y)$ in the bulk. This expression describes a monotonically
decreasing function that interpolates between the two boundary values $a_{0}$
and $a_{L}$.

The remaining nontrivial condition which
relates $\hat{\phi}_{L}$ to $\hat{\phi}_{0}$ is the continuity of the $\hat{
\phi}$ field in the bulk,
\begin{equation}
\hat{\phi}_{L}=\hat{\phi}_{0}+\int_{0}^{L}\hat{\phi}^{\prime }(y)\,dy=
\hat{\phi}_{0}+c\int_{0}^{L}\frac{a_{0}^{4}}{a^{4}(y)}\,dy\,.  \label{contphi}
\end{equation}
This equation, together with the boundary conditions and explicit forms for 
$a^{4}$ and $\hat{\phi}^{\prime }$, lead to an {\em overdetermined }set of
algebraic equations. This means that, in general, no static solution can be
found unless one extra fine tuning on the original parameters, $\Lambda _{1},
$ $\Lambda _{2},$ $V_{1}$ and $V_{2}$, is imposed. We note that the
form of the interaction terms of the scalar field $\hat{\phi}$ on the
two branes completely determines the ratio of the values of the
conformal factor on the two branes. More specifically, 
\begin{equation}
\biggl(\frac{a_{0}}{a_{L}}\biggr)^{4}=\frac{\sinh (\omega y_{0})}{\sinh
[\omega \,(y_{0}-L)]}=-\frac{(\partial _{\hat{\phi}}V_{2})_{y=L}}{(\partial
_{\hat{\phi}}V_{1})_{y=0}}\,,  \label{cond}
\end{equation}
from which we further conclude that the derivatives of the interaction terms
on the branes with respect to $\hat{\phi}$ should have opposite signs in
order to achieve static solutions. The above relation also leads to the
determination of the distance $L$ between the two branes in terms of the
fundamental parameters of the theory.  In the limit of
large $|\Lambda_B|$ and $\omega y_{0},\omega \,(y_{0}-L)\gg 1$,
eq. (\ref{cond}) is simplified and leads to the result 
\begin{equation}
L=\sqrt{\frac{3}{8\hat{\kappa}^{2}|\Lambda _{B}|}}\,\ln\,\Biggl|\frac{
(\partial _{\hat{\phi}}V_{2})_{y=L}}{(\partial _{\hat{\phi}}V_{1})_{y=0}}
\Biggr|\,,
\end{equation}
which resembles the one derived by Goldberger and Wise \cite{GW}. In that
case, the distance between the two branes remains fixed as long as the
bulk scalar field assumes different vacuum expectation values on the
two branes. As we can see from the example above, the distance between
the branes is completely  determined by the requirement of the
time independence of the metric,  equivalent to the cancellation of the
effective cosmological constant.   This conclusion is quite generic and
holds for arbitrary interaction terms.  Thus, in the case of the
massless scalar, the fixed inter-brane distance is  the consequence of
the fine-tuning of cosmological constant rather than a true dynamical
stabilization. Similarly, in the limit of  small cosmological
constant and small $\omega\,y_{0}$ and  $\omega \,(y_{0}-L)$, $a^{4}(y)$
becomes a linear function of $y$ and the distance between the two branes
is given by the expression 
\begin{equation}
L=y_{0}\,\biggl(1-\Biggl|\frac{(\partial _{\hat{\phi}}V_{1})_{y=0}}{
(\partial _{\hat{\phi}}V_{2})_{y=L}}\Biggr|\biggr)=\sqrt{\frac{3}{\hat{\kappa
}^{2}}}\,\biggl(\frac{1}{|(\partial _{\hat{\phi}}V_{1})_{y=0}|}-\frac{1}{
|(\partial _{\hat{\phi}}V_{2})_{y=L}|}\biggr)\,.
\end{equation}
In the above, we have used the definitions (\ref{def}) and the jump condition
of the scalar field on the two branes. Once again, the distance between
the two branes is uniquely determined and the derivatives of the interaction
terms of the bulk scalar field on the branes should be different.

As an illuminating example, we consider the case of linear interaction
terms, i.e. $V_1(\hat \phi)=\alpha\,\hat\phi$ and $V_2(\hat
\phi)=\beta\,\hat\phi$. According to eq. (\ref{cond}), the coefficients
$\alpha$ and $\beta$ should be chosen in such a way as to satisfy
$\alpha\,\beta <0$. This statement is rather important since, unless the two
branes have opposite ``charges" with respect to $\phi$, no static solutions
arise in the above framework. Then, the expression for the distance $L$
between the two branes is simplified and is found to be 
\begin{equation}
L=\sqrt{\frac{3}{8 \hat \kappa^2 |\Lambda_B|}}\,\ln\,\biggl|\frac{\beta}{
\alpha}\biggr|\,,
\end{equation}
in the case of a large bulk cosmological constant, while in the opposite
case, we obtain 
\begin{equation}
L=\sqrt{\frac{3}{\hat \kappa^2}}\,\biggl(\frac{1}{|\alpha|}- \frac{1}{|\beta|
}\biggr)\,.
\end{equation}

We now turn to the jump conditions that the solution for the conformal
factor must satisfy on the two branes. Working again in the limit of large 
$\omega y_{0}$ and $\omega \,(y_{0}-L)$ and rearranging the jump conditions
for $a(y)$ that appear in eqs. (\ref{jump1})-(\ref{jump2}), we obtain the
following conditions 
\begin{equation}
\sqrt{\frac{6\,|\Lambda _{B}|}{\hat{\kappa}^{2}}}=-\Bigl[\Lambda _{2}+V_{2}(
\hat{\phi}_{L})\Bigr]=\Lambda _{1}+V_{1}(\hat{\phi}_{0})\,.  \label{con1}
\end{equation}
The only remaining free parameter is the value of the scalar field on one of
the two branes as $\hat{\phi}_{L}$ and $\hat{\phi}_{0}$ are related as follows
\begin{equation}
\hat{\phi}_{L}\simeq \hat{\phi}_{0}+\frac{c}{\omega}\exp [-\omega
(y_{0}-L)].
\end{equation}
Similarly, in the limit of small $\omega y_{0}$ and $\omega \,(y_{0}-L)
$ we obtain the relations
\begin{equation}
\frac{3}{2\hat{\kappa}^{2}}=-\Bigl[\Lambda _{2}+V_{2}(\hat{\phi}_{L})\Bigr
]\,(y_{0}-L)=\Bigl[\Lambda _{1}+V_{1}(\hat{\phi}_{0})\Bigr]\,y_{0}\,,
\label{con2}
\end{equation}
where the values of the scalar fields on the two branes are related
by 
\begin{equation}
\hat{\phi}_{L}=\hat{\phi}_{0}+cy_0\ln \frac{
y_{0}}{y_{0}-L}\,.
\end{equation}

By choosing, for example, $\hat{\phi}_{0}$ to satisfy the condition $\Lambda
_{1}+V_{1}(\hat{\phi}_{0})\,=\Lambda _{RS}$ in eq. (\ref{con1}), we are
left with one fine tuning imposed on some combination of the fundamental
parameters of the theory. The above result leads to the conclusion that,
despite the presence of the bulk scalar field, the stabilization of the
extra dimension still relies on the correlation that holds between the
energy densities of the two branes and the bulk cosmological constant. In
the limit $V_{1}$, $V_{2}\rightarrow 0$, we recover the condition that holds
between the bulk and brane cosmological constants in the case of the
Randall-Sundrum model \cite{RS}. In that case, every distance $L$ between
the two branes is acceptable as long as the correlation between the energy
densities of the two branes holds. In our case, for non-vanishing
$V_{1}$ and $V_{2}$, a unique value of the distance $L$ emerges which
is mainly determined by the first derivatives of the interaction terms
with respect to 
$\hat{\phi}$. However, once the interaction terms have been chosen, the
consistency of the solution, and thus the viability of the whole scenario,
relies on the careful choice of the two self-energies, $\Lambda _{1}$ and
$\Lambda _{2}$, in such a way as to satisfy the constraint (\ref{con1}).
Alternatively, for fixed $\Lambda_i$, one must fine tune the parameters
of the interaction terms $V_{i}$ according to eq. (\ref{con1}) and this 
fine-tuning will change the distance between the two branes.

 Another solution of the differential equation (\ref{basic})
given in terms of $cosh(\omega |y-y_{0}|)$ was rejected being inconsistent
with the original equations (\ref{fin1}) and (\ref{fin2}). Such a solution,
if acceptable, would describe a conformal factor characterized by the
existence of a minimum at $y=y_{0}$ with both branches going upwards as one
approaches the two branes. Mathematically, this solution would be consistent
with the equations of motion only if the sign of the kinetic term of the
bulk scalar field in eq. (\ref{action}) were exactly the opposite. If we
treat this case formally, both of the energy densities of the branes could
be positive, however, the correlation between these two would still
remain.The ``wrong'' sign of the kinetic term for the bulk field would
correspond to a tachyonic mode and signal an intrinsic
instability of such a construction. The appearance of tachyonic modes, in
the absence of a monotonic configuration of the bulk scalar field along the
extra dimension, was also pointed out in~\cite{tanaka}.

It appears that the only solution without a fine tuning between the brane
self-energies and the bulk cosmological constant is the single-brane
configuration with the extra dimension ending on a singularity. Indeed,
going back to the solution for the conformal factor $a(y)$,
eq. (\ref{solneg}), and the first derivative of the bulk scalar field,
eq. (\ref{scalar}), we observe that the former quantity vanishes, while
the latter diverges, at $y=y_{0}$.  By evaluating the scalar curvature
$R$, which is given by the expression
\be
R=\omega^2\,\biggl[\,\frac 34\, \coth^2(\omega|y-y_0|)
-2 \biggr]\,,
\ee
one can easily check that a true spacetime singularity occurs at the 
point $y=y_0$. The solution for the conformal factor is still  given by eq.
(\ref{solneg}) while the size of the extra dimension is set by the
position of the singular point which can be found from the boundary
conditions for $\hat{\phi}^{\prime}$. In the case of a linear
interaction of the bulk scalar field with the single brane, 
$V(\hat\phi)=\alpha\,\hat \phi$, the position of the singular point is
given by
\begin{equation}
y_{0}=\sqrt{\frac{3}{8\hat{\kappa}^2|\Lambda_B|}}\,Arc\,\sinh\,
\frac{4\sqrt{|\Lambda _{B}|}}{|\alpha |}\,.
\end{equation}
The boundary condition for the scale factor can be satisfied
by the appropriate choice of $\hat{\phi}_{0}$. By performing an analysis
similar to that of Ref. \cite{ck}, we can easily see that
the conservation of energy and momentum is not violated near the
singularity for any massless particle, as well as for any massive
excitations independent of $y$, propagating in the given spacetime
background. In the limit of
small cosmological constant in the bulk, $|\Lambda _{B}|\ll \alpha ^{2}$,
the distance to the singular point is inversely proportional to the size of
the coupling $\alpha $. In the opposite limit of small coupling constant,
$|\Lambda _{B}|\gg \alpha ^{2}$, the solution for the scale factor is simply
a falling exponent everywhere apart from the small $\omega ^{-1}$ vicinity
of $y_{0}.$ Thus, in this limit, this solution is basically the one found by
Randall and Sundrum \cite{RS2} with the exponential tail being cut off at
the finite distance $y_{0}.$ For a vanishing value of the coupling,
$\alpha
\rightarrow 0,$ this point is at infinity, $y_{0}\rightarrow \infty$, the
correlation between the brane self-energy and the bulk cosmological constant
reappears and the solution coincides exactly with that in Ref. \cite{RS2}.
The presence of a singularity was recently advocated to solve the
hierarchy problem \cite{ck} and the cosmological constant problem
\cite{cc}. 

However, the absence of a second brane at $y<y_0$ does not mean that 
the system is not overdetermined. An accurate consideration of the 
singularity suggests that the consistency of the solution requires  
fixing boundary conditions for the conformal factor and the scalar field 
which is equivalent to assuming certain source terms at the 
singularity \cite{Nilles}. In order to have a consistent treatment of
the boundary conditions at the singularity, it is helpful to return 
to our two-brane solution and consider the limit 
$\Lambda_2\rightarrow -\infty$. The boundary condition for the scalar
field 
requires that $|(\partial _{\hat{\phi}}V_{2})_{y=L}|\rightarrow \infty$,
as it 
can be easily seen from eq. (\ref{cond}). It can be further shown that 
the two limits have to be taken in a {\em correlated} way, in order to
fulfill the condition (\ref{con2}). This condition shows explicitly that 
the boundary conditions at the singularity are correlated with $\Lambda_1$
and $V_1$.

\section{Conclusions}

It is well established that the stabilization of
an extra dimension, by the introduction of a stabilizing potential for
the radion field, leads to the resolution of several cosmological
paradoxes and to the restoration of the standard Friedmann equation on
our brane-universe 
\cite{kkop,csak2,kkop2,kim}. The
stabilizing potential produces a non-vanishing (55)-component for the
energy-momentum tensor which is proportional to the trace of the
energy-momentum tensor of our brane. When a ``phenomenological''
potential  for the size of the extra dimension is introduced, this
adjustment of 
$T_{55}$ to the required value is automatic \cite{kkop2}. 

Here, we have shown that it is possible to
stabilize two positive self-energy branes. The ratio
of the scale factors on these two branes is determined through the relative
detuning of brane self-energies from the limiting value $\sqrt{6|\Lambda
_{B}|/\hat{\kappa}^{2}}.$ The time independence of this solution, equivalent
to the cancellation of the effective cosmological constant, comes as an extra
condition to which the stabilization mechanism must satisfy. When the
brane self-energies are specified, this condition determines uniquely the
distance between the two branes. 

Next, we considered a candidate mechanism for a dynamical 
stabilization of the extra dimension. We
introduced a bulk scalar field with arbitrary interaction terms on
the two branes. By choosing a vanishing potential for the scalar field
in the bulk, the exact solution for the conformal factor, in the
presence of the scalar field, was determined for zero, positive and
negative bulk cosmological constant. In all cases, the solution for
$a(y)$ was accompanied by the appearance of a true spacetime
singularity at a finite point $y_{0}$ along the extra dimension. The
singularity could only be avoided by placing the second brane at a
distance $L<y_{0}$.  It
was shown that the ratio of the values of the first derivatives, with
respect to 
$\hat{\phi}$, of the interaction terms on the two branes completely
determines the ratio of the boundary values of the conformal factor and,
moreover, the distance between the two branes. This, according to our
analysis, is a generic result independent of the form of the interaction
terms of the scalar field on the two branes. However, it would be wrong
to conclude that the introduction of a scalar field in the bulk may,
indeed, leads to the desired stabilization of the inter-brane distance.
In some sense, the fixed size of the extra dimension is the result of
{\em imposing} the time independence of the solution which lead to
the overdetermined set of equations. Indeed, we were able to
demonstrate that the above result is always accompanied by the need for
the correlation of the self-energies of the two branes and the coupling
constants of the scalar field with the branes. As in the case of the
original Randall-Sundrum model \cite{RS}, the static solution exists
only if the total energy density of one of the two branes (now, given
by the sum of the brane self-energy and the scalar interaction term) is
negative. This result mars the significance of the successful
stabilization of the extra dimension as it introduces an unphysical,
and phenomenologically unacceptable, assumption.

It appears that the only solution where the unphysical correlation is not
required is the single-brane configuration with the extra dimension ending
on the true singularity. The position of this singularity can be
interpreted
as the size of the extra dimension and depends on the size of the scalar
field-brane coupling constant. For a small value of the coupling, this
solution generalizes a single brane/infinite dimension configuration
discussed in \cite {RS2} and shows that the presence of a scalar field in
the bulk leads to the cutoff of the exponential fall of the scale factor.
This static solution could be important as it  represents an example,
where the
effects of presumably  large bulk cosmological constant and brane
self-energy are  completely screened by a massless scalar. Similar
observations were made  recently in \cite{cc}. Unfortunately, 
the correct way of treating the singularity \cite{Nilles} requires the 
explicit fixing of boundary conditions for the scalar field and metric at
the singular point, which reinstates the fine-tuning problem observed in 
this work for the two-brane model.



\end{document}